\documentclass{PoS}

\title{Applications of KP Nuclear Parton Distributions}

\ShortTitle{KP Nuclear PDFs}

\author{S. A. Kulagin\\
        Institute for Nuclear Research of the Russian Academy of Sciences, Moscow 117312, Russia\\
        E-mail: \email{kulagin@ms2.inr.ac.ru}}

\author{\speaker{R. Petti}\\
        Department of Physics and Astronomy, University of South Carolina, Columbia SC 29208, USA\\
        E-mail: \email{roberto.petti@cern.ch}}

\abstract{%
We review the nuclear parton distribution functions computed on the basis of
our microscopic model taking into account a number of nuclear effects including
Fermi motion and nuclear binding, nuclear meson-exchange currents, off-shell corrections
to bound nucleon distributions and nuclear shadowing.
We discuss applications to a number of different processes including  
lepton-nucleus deep inelastic scattering, proton-nucleus Drell-Yan lepton pair production at Fermilab, 
as well as $W^\pm$ and $Z^0$ boson production in proton-lead collisions at the LHC.   
}

\FullConference{XXIV International Workshop on Deep-Inelastic Scattering\\
                11--15 April 2016\\
                DESY Hamburg, Germany}

\begin{document}

\section{Introduction}
\label{sec:intro}

The parton distribution functions (PDFs) are universal process-independent characteristics
of the target at high invariant momentum transfer $Q$, which are driven by non-perturbative
strong interactions in the considered target.
The QCD factorization theorem~\cite{Collins:1989gx}, which is expected to hold for 
both nucleons and nuclei, allows to use the same PDFs in different hard processes 
involving lepton and hadron probes.
In the context of nuclear targets, nuclear PDFs (NPDF) are usually extracted from data using phenomenological parameterizations of the dependencies of nuclear corrections on 
both the Bjorken $x$ and the atomic number $A$. Although these QCD-based studies are useful in constraining nuclear effects for different partons, they provide limited information about the underlying physics mechanisms responsible of the nuclear modifications of PDFs.

In this contribution we briefly review a different approach to NPDFs~\cite{KP04, KP14}, 
which computes nuclear corrections on the basis of a microscopic model
incorporating several nuclear physics mechanisms, including the smearing with the 
energy-momentum distribution of bound nucleons (Fermi motion and binding), 
the off-shell correction to bound nucleon PDFs, the contributions from meson 
exchange currents and the coherent propagation of the hadronic component
of the virtual intermediate boson in the nuclear environment.
This model explains to a high accuracy the observed
$x$, $Q$ and $A$ dependencies of the measured nuclear effects in
deep-inelastic scattering (DIS)
on a wide range of targets from deuterium to lead~\cite{KP04,KP07,KP10,AKP16};
the magnitude, the $x$ and mass dependence of available data from Drell-Yan (DY) 
production off various nuclear targets~\cite{KP14}; as well as the differential cross-sections
and asymmetries of the $W^\pm$ and $Z$ boson production in p+Pb collisions at the LHC~\cite{Ru:2016wfx}.

\section{Model for Nuclear Parton Distributions}
\label{sec:model}

In the model of Ref.\cite{KP04,KP14} the nuclear PDF $q_{a/A}$ of flavor $a$ in the 
nucleus $A$ receives a number of different contributions, which can be 
summarized as 
follows~\footnote{For brevity we suppress the explicit dependencies on $x$ and $Q^2$.}:
\begin{equation}
\label{npdf}
q_{a/A} = \left\langle q_{a/p}\left(1+\delta f\frac{p^2-M^2}{M^2}\right)\right\rangle +
		\left\langle q_{a/n}\left(1+\delta f\frac{p^2-M^2}{M^2}\right)\right\rangle
          + \delta q_a^\mathrm{MEC} + \delta q_a^\mathrm{coh} .
\end{equation}
The first two terms on the right side of the equation represent the contribution from the
incoherent scattering off bound protons and neutrons, while the 
terms $\delta q_a^\mathrm{MEC}$ and $\delta q_a^\mathrm{coh}$ are the corrections
arising from nuclear meson exchange currents (MEC) and
coherent interactions of the intermediate virtual boson with the nuclear target, respectively.
Since PDFs are universal Lorentz-invariant functions, although Eq.(\ref{npdf}) 
is written for DIS in the nucleus rest frame, results can be used to 
describe different processes in any reference frame.

The brackets appearing in the first two terms of Eq.(\ref{npdf}) imply the averaging 
(convolution) of the proton (neutron) PDF with the proton (neutron) spectral function
describing the energy-momentum distribution of bound nucleons%
~\cite{Kulagin:1989mu,Kulagin:1994fz,KP04,KP14}.
Since the bound nucleons are, in general, off their mass-shell, the corresponding PDFs
explicitly depend on the nucleon invariant mass squared $p^2$.
In the vicinity of the mass shell, the off-shell correction (OS) linearly depends on the 
virtuality $p^2-M^2$ and the function $\delta f = \partial\ln q_{p}/\partial\ln p^2$ describes the
relative off-shell modification of nucleon PDFs.
This special nucleon structure function does not contribute to the cross section of the
physical nucleon, but it is relevant only for the bound nucleon and describes its
response to the interaction in a nucleus.

The mesonic fields mediate the nucleon-nucleon interaction at distances exceeding
the typical nucleon size and also contribute to the quark-gluon content of the nucleus.
The term $\delta q_a^\mathrm{MEC}$ in Eq.(\ref{npdf}) is the contribution from the
virtual mesons exchanged (MEC) between bound nucleons.
This contribution is important in order to balance the overall
nuclear light-cone momentum and is constrained by the light-cone momentum sum rule,
similarly to the case of the gluon contributions at the partonic level.

The last term in Eq.(\ref{npdf}) is due to the propagation effects of the intermediate hadronic
states of the virtual boson in the nuclear environment.
The term $\delta q_a^\mathrm{coh}$ is relevant at low $x$ and
involves contributions from the Glauber-Gribov multiple scattering series
describing the interaction of the intermediate hadronic states with bound nucleons.
At small $x$ this correction is negative, giving rise to the nuclear shadowing (NS) effect.
We also note that in the transition region $x> 0.05$ this correction may be positive
for the nuclear valence quark distributions because of a constructive interference 
between the scattering amplitudes in the $C$-even and $C$-odd channels~\cite{KP04,KP14}.

\begin{figure}[t]
\centering
\includegraphics[width=\textwidth]{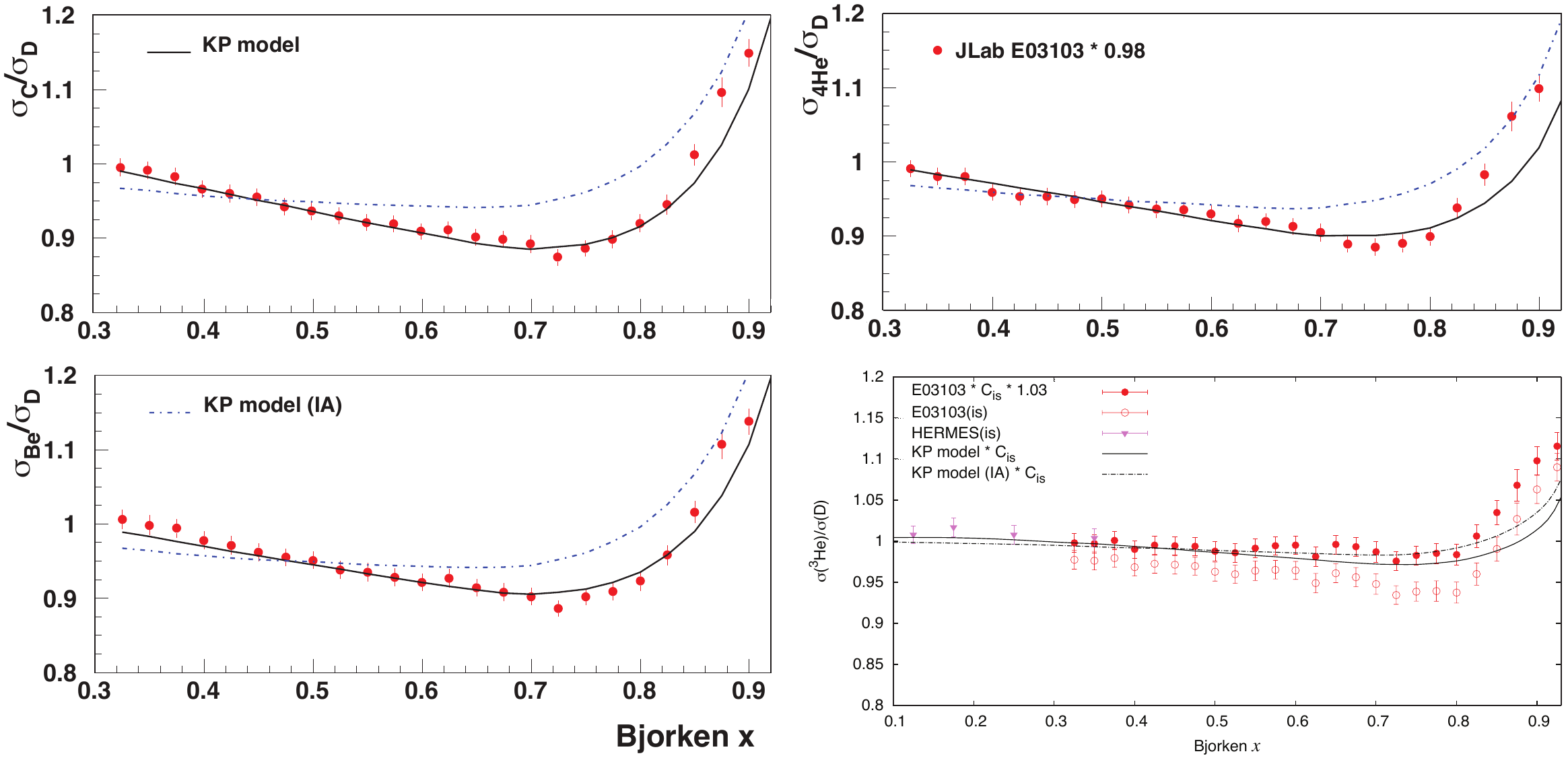}
\caption{%
Ratios of DIS cross-sections of ${}^{12}$C (top left), ${}^{9}$Be (bottom left),
${}^{4}$He (top right), and ${}^{3}$He (bottom right)
with respect to deuterium compared with our predictions for the same kinematics.
Data points are from the E03-103 experiment at JLab~\cite{Seely:2009gt} (points with
$x>0.85$ have $W^2<2$ GeV$^2$) and the HERMES experiment at HERA~\cite{Ackerstaff:1999ac},
with statistical and systematic uncertainties added in quadrature.
The result of a calculation in the impulse approximation with no off-shell correction is also shown as
dashed-dotted line for comparison.
Following the analysis of Ref.\cite{KP10} a common normalization factor of 0.98 (1.03 for ${}^{3}$He)
is applied to the data from Ref.\cite{Seely:2009gt} for consistency with SLAC and NMC data.
For more details see Ref.\cite{KP10}.
\label{fig:DIS-E03013}}
\end{figure}

We note that nuclear effects in different kinematical regions of
$x$ are related by the DIS sum rules and normalization constraints.
In the model of Ref.\cite{KP04,KP14} these conditions serve as 
dynamical constraints on $\delta q_a^\mathrm{coh}$, allowing to calculate the coherent
nuclear corrections for different combinations of nuclear PDFs in terms of the 
off-shell function $\delta f$ and the bound nucleon virtuality.

\section{Results and Discussion}
\label{res}	

A thorough analysis of data on the ratios of DIS structure functions off different nuclei ranging from
${}^4$He to ${}^{208}$Pb was carried out in Ref.\cite{KP04}. The model demonstrated an excellent
performance and was able to describe the observed $x$, $Q^2$ and $A$ dependencies of data to a high accuracy.
The predictions of Ref.\cite{KP04} were verified~\cite{KP10} with the more recent
nuclear DIS data from
the E03-103 experiment at JLab~\cite{Seely:2009gt} and the HERMES experiment at HERA~\cite{Ackerstaff:1999ac}.
Figure~\ref{fig:DIS-E03013} shows a comparison of our predictions with the data from the recent
measurement of Ref.\cite{Seely:2009gt} for light nuclei $^{12}$C, $^9$Be, $^4$He and $^3$He,
focused at intermediate and large Bjorken $x$.
In this region the relevant nuclear corrections are due to
the Fermi-motion and nuclear binding correction~\cite{FMB,Kulagin:1989mu,Kulagin:1994fz,KP04,KP14},
as well as to the off-shell correction~\cite{Kulagin:1994fz,KP04}.
The dashed curve in Fig.\ref{fig:DIS-E03013} shows the result obtained in the 
impulse approximation with no off-shell correction, while the solid line represent the 
predictions from the full model. This comparison illustrates the relevance of the OS correction,
which, together with the Fermi-motion and nuclear binding correction,
provides a quantitative description of the observed EMC effect. An independent study from 
DIS off proton and deuteron targets confirms the universality of the $\delta f$ function~\cite{AKP16}. 
We also note that high-twist corrections~\cite{Alekhin:2007fh}
as well as target-mass corrections
play a significant role in the results of Refs.\cite{KP04,KP10,AKP16}.

\begin{figure}[t]
\centering
\includegraphics[width=0.75\textwidth,height=0.305\textheight]{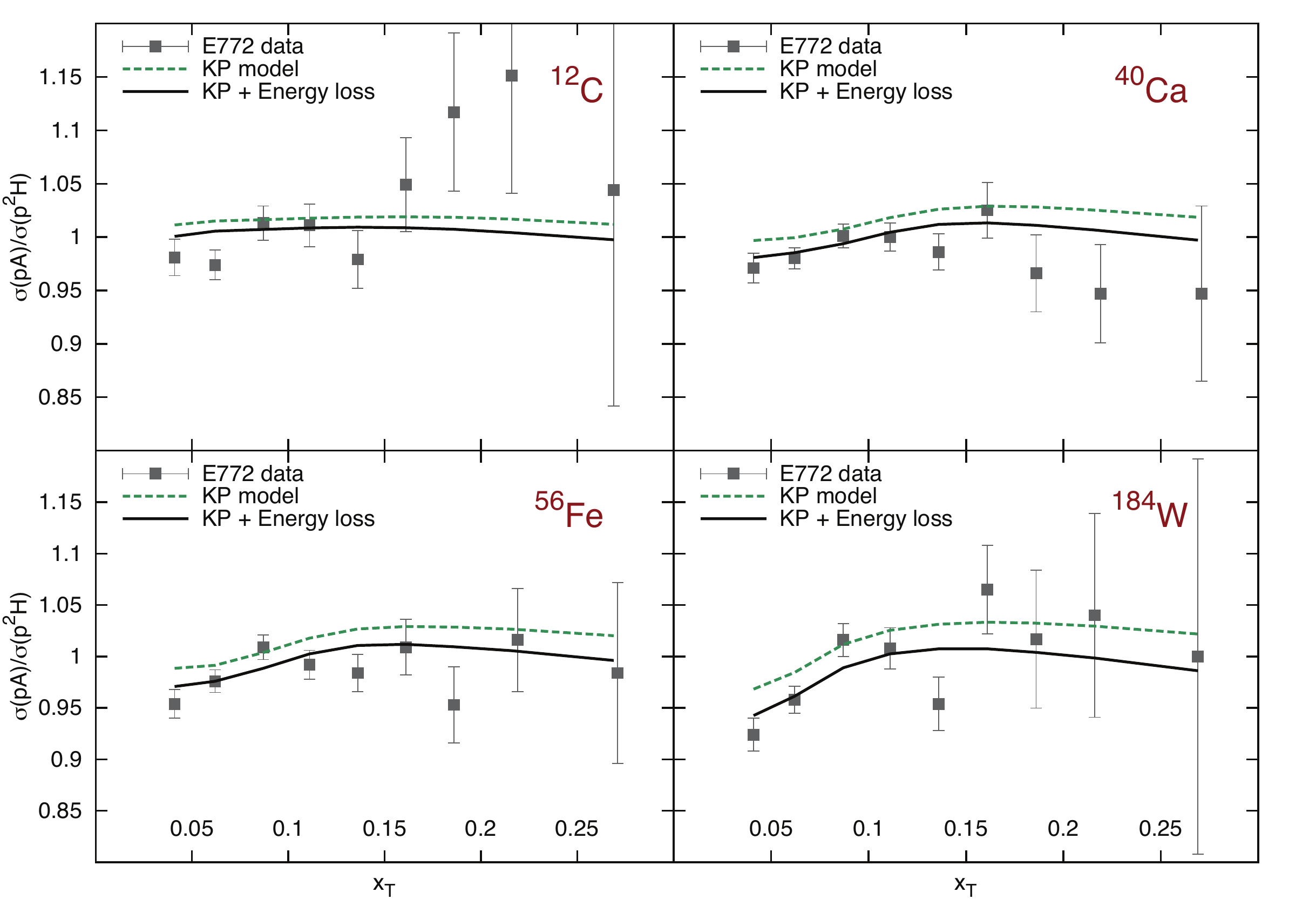}
\caption{%
Ratios of the DY cross sections of ${}^{12}$C (top left), ${}^{40}$Ca (top right),
${}^{56}$Fe (bottom left), and ${}^{184}$W (bottom right)
with respect to deuterium as a function of the
Bjorken $x_T$ of the partons in the target nucleus.
Data points are from the E772 experiment~\cite{E772}.
The curves are the predictions of Ref.\cite{KP14}
with (solid) and without (dashed) the energy loss correction to the projectile quark. 
For more details see Ref.~\cite{KP14}. 
\label{fig:DY-E772}}
\end{figure}

The model of Ref.\cite{KP14} predicts that nuclear corrections are different for the 
nuclear valence and sea distributions, and that they also depend on the PDF flavor
(see also Ref.\cite{Kulagin:2015lkm}).
The Drell-Yan process (DY)  can be particularly relevant in this context.
For the kinematics of the E772 and E866 experiments \cite{E772,E866}
the ratio of the DY cross sections in the region of the Bjorken variable of the nuclear 
target $x_T < 0.15$ is driven by the antiquark distributions in the target nucleus.
The predictions of the KP model were  compared with the available nuclear 
DY data \cite{E772,E866} in Ref.\cite{KP14}.
One interesting feature of these data is that the DY cross-section ratios 
do not show any significant \emph{antishadowing} enhancement at $x_T\sim 0.1$ \cite{E772}.
This behavior has been a long standing puzzle, since the nuclear
binding should result in an excess of nuclear mesons,
which is expected to produce a marked enhancement in the nuclear
anti-quark distributions~\cite{Bickerstaff:1985ax}.
Our model predicts a significant cancellation of
different nuclear effects for the antiquark distributions in the region $x_T\sim 0.1-0.15$.
This effect is in a good agreement with the available nuclear DY data,
as illustrated in Fig.\ref{fig:DY-E772} for the E772 experiment.
We also note that the nuclear dependence of the DY process can be affected by the 
initial state interaction of the projectile particle (parton) within the nuclear environment.
In our study we treat this effect phenomenologically, assuming a modification of the 
projectile Bjorken variable $x_B$ due to quark energy loss effect~\cite{Garvey:2002sn}.
We found that the data from the E772 and E866 experiments are consistent
with the presence of moderate energy loss effects, with a rate of the order of 
1\,GeV/fm.

\begin{figure}[t]
\centering
\includegraphics[width=\textwidth]{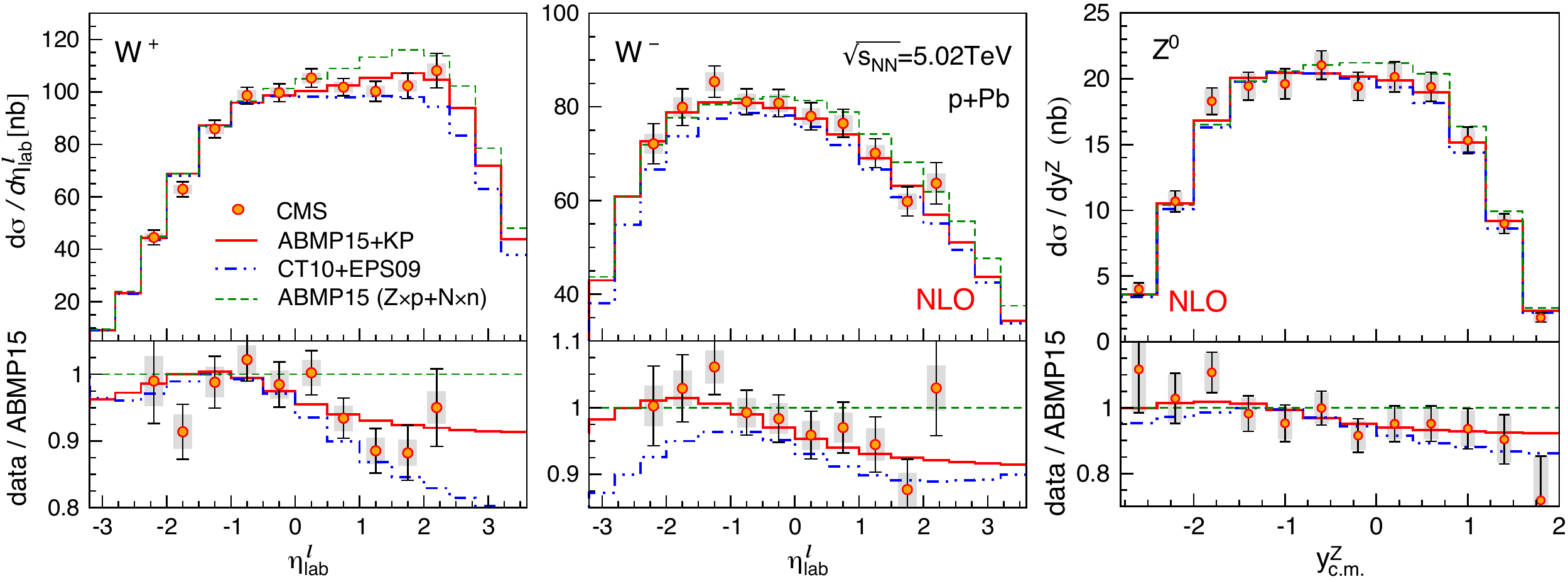}
\caption{%
Top panels: differential cross-sections for $W^+$ (left), $W^-$ (middle), and $Z^0$ (right) production
in p+Pb collisions at $\sqrt{s_{NN}}=5.02$~TeV, as a function of (pseudo)rapidity.
The data points indicate the CMS measurements~\cite{Khachatryan:2015hha,Khachatryan:2015pzs},
while the curves show the predictions based on different models: ABMP15+KP (solid),
CT10+EPS09 (dashed-dotted), and ABMP15 without nuclear modifications (dashed).
Bottom panels: ratios of the data points and the model predictions shown in the top panels with respect
to the result obtained without nuclear modifications (ABMP15).
For more details see Ref.\cite{Ru:2016wfx}.
\label{fig:WZ-CMS}}
\end{figure}

A study of the $W^\pm$ and $Z^0$ boson production cross sections
in p+Pb collisions with $\sqrt{s}=5.02$\,TeV at the LHC was performed in Ref.\cite{Ru:2016wfx} 
in terms of the NPDF model of Ref.\cite{KP14}.
A detailed comparison with the recent precision data from the CMS and ATLAS experiments at the LHC
clearly indicates the presence of nuclear modifications on the $W/Z$ boson production cross sections
with respect to the case of p+p collisions.
We found an excellent agreement between the theoretical predictions based on KP NPDF
and the measured observables in the entire kinematic range accessible by the experiments,
as illustrated in Fig.\ref{fig:WZ-CMS}.
In particular, the model correctly describes the flavor dependence of the nuclear modifications
observed in both the $W^+$ and $W^-$ boson distributions.
The full nuclear corrections
on the $W/Z$ boson production in p+Pb collisions are the result of an interplay of the 
various physics mechanism discussed in Sec.\ref{sec:model}.
Finally, it is worth noting that the precision currently achieved by the LHC experiments -- most
notably with the latest CMS measurements of $W/Z$ boson production -- starts to be sensitive to
the predicted nuclear corrections. A further improvement of the accuracy of future data sets
would be extremely valuable in this context since it could allow to disentangle the effect of
different underlying mechanisms responsible for the nuclear modifications of PDFs and to
study their flavor dependence.


The work of S.K. was supported by the Russian Science Foundation grant No.~14-22-00161.
R.P. was supported by the grant DE-SC0010073 from the Department of Energy, USA.

\end{document}